\documentclass[twocolumn,color,superscriptaddress,psfig,showpacs,amsmath,amssymb,floatfix,fleqn]{revtex4-1}
\usepackage{epsf}

\usepackage{amssymb}

\usepackage{natbib}
\setcitestyle{square,numbers}
\usepackage{amsthm}
\usepackage{graphicx}
\usepackage{dcolumn}
\usepackage{bm}
\usepackage{color}
\usepackage{subfigure}
\usepackage[subnum]{cases}
\usepackage{float}
\usepackage{epstopdf}
\usepackage{tabularx}
\usepackage{multirow}
\usepackage{setspace}

\begin{document}


\title{Radio-wave communication with chaos}

\author{Hai-Peng Ren}\thanks{Corresponding author:Hai-Peng Ren,renhaipeng@xaut.edu.cn}
\affiliation{Shaanxi Key Laboratory of Complex System Control and Intelligent Information Processing, Xi'an University of Technology, Xi'an 710048, China}

\author{Wu-Yun Zheng}
\affiliation{Shaanxi Key Laboratory of Complex System Control and Intelligent Information Processing, Xi'an University of Technology, Xi'an 710048, China}

\author{Celso Grebogi}
\affiliation{Shaanxi Key Laboratory of Complex System Control and Intelligent Information Processing, Xi'an University of Technology, Xi'an 710048, China}
\affiliation{Institute for Complex System and Mathematical Biology, SUPA, University of Aberdeen, Aberdeen AB243UE, United Kingdom}

\date{\today}

\begin{abstract}
The invariance of the Lyapunov exponent of a chaotic signal as it propagates along a wireless transmission channel provides a theoretical base for the application of chaos in wireless communication. In additive Gaussian channel, the chaotic signal is proved to be the optimal coherent communication waveform in the sense of using the very simple matched filter to obtain the maximum signal-to-noise ratio. The properties of chaos can be used to reduce simply and effectively the Inter-Symbol Interference (ISI) and to achieve low bit error rate in the wireless communication system. However, chaotic signals need very wide bandwidth to be transmitted in the practical channel, which is difficult for the practical transducer or antenna to convert such a broad band signal. To solve this problem, in this work, the chaotic signal is applied to a radio-wave communication system, and the corresponding coding and decoding algorithms are proposed. A hybrid chaotic system is used as the pulse-shaping filter to obtain the baseband signal, and the corresponding matched filter is used at the receiver, instead of the conventional low-pass filter, to maximize the signal-to-noise ratio. At the same time, the symbol judgment threshold determined by the chaos property is used to reduce the Inter-Symbol Interference (ISI) effect. Simulations and virtual channel experiments show that the radio-wave communication system using chaos obtains lower bit error rate in the multi-path transmission channel compared with the traditional radio-wave communication system using Binary Phase Shift Keying (BPSK) modulation technology and channel equalization.
\end{abstract}

\maketitle


\section{\label{Intro}Introduction}
Chaos, having a random like deterministic dynamics, has many properties suitable for communication applications, such as wide spectrum, sensitivity to initial data and pulse-like auto-correlation. The realization of chaos control \cite{PhysRevLett.64.1196}  and synchronization \cite{PhysRevLett.64.821} has laid the foundation for the application of chaotic signals. Hayes \emph{et al}. used perturbation control to encode information sequence into chaotic signals that were used in communication \cite{PhysRevLett.70.3031}. Since then, many communication methods have been proposed, such as Chaos Masking \cite{PhysRevLett.71.65,Milanovic2002Improved}, Chaos Modulation \cite{Abel2002Chaos,Yang1996Secure}, Chaos Spread Spectrum \cite{Shake2005Security}, Chaos Shift Keying \cite{Dedieu1993Chaos} and their improvements \cite{Kennedy2000Performance,Zhang2015A,Hu2017Design}.

The research of chaotic communication was originally focused on the security performance in ideal channels with white noise. With further developments, the research hotspot has gradually turned to the use of chaos to improve the performance of traditional communication systems. Since Argyris \emph{et al}. used chaotic signals to improve the bit transmission rate in the optical fiber communication system \cite{Argyris2005Chaos}, the research on communication with chaos shifted from ideal channels to the practical communication channels. Compared with the ideal channel, the wireless channel is a practical channel suffering from more complex constraints, such as multi-path propagation, Doppler shift, \emph{etc} \cite{Williams2001radiochannels}. Whether chaos could be applied to wireless channel became a basic problem. Recent work showed that the information in chaotic signal is preserved after transmitted through wireless channel \cite{Ren2013Wireless}, at the same time, experimental circuit encoding and decoding method for wireless communication was designed to validate the theoretical predictions \cite{Ren2016Experimental}. Besides these advances, other advantages for using chaos into communication have been reported recently, including the use of simple matched filter to achieve the maximum signal to noise ratio \cite{Corron2016Analytically}, and a decoding threshold using chaos property to relief ISI caused by multi-path propagation \cite{Yao2017Chaos}. All these progresses pushed the research forward to the practical application of chaos in communication systems. However, it is very difficult for the current available transducers or antennas to convert the chaotic signal into the media waveform (radio-wave or acoustic signal \emph{et al}.), because the chaotic signal possesses wide spectrum bandwidth, and the transducers or antennas and the associate amplifiers have relative narrow bandwidth. This fact prevents the chaos application in practical communication systems. The coping with this obstacle is the key for future research work on communication with chaos. In this paper, we show that the simplest way to solve the above obstacle is by using a chaotic pulse-shaping filter to replace the conventional pulse-shaping filter. Using radio-wave communication system as example, we compare the performance of the radio wave communication system using the chaotic pulse-shaping filter and corresponding matched filter with the conventional radio wave communication system. In such a way, we use the merits that chaotic signal provides to improve the practical performance of conventional communication system.\\
\begin{figure*}
\includegraphics[scale=0.7,trim= 0 0 0 0] {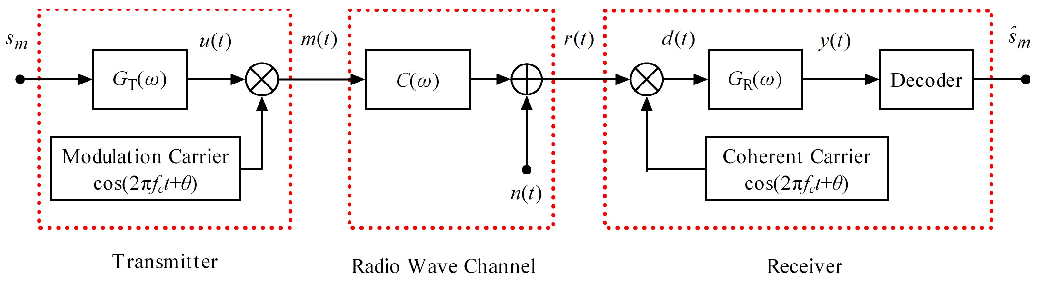}
\caption{\label{fig.1} Block diagram of radio wave communication system.}
\end{figure*}
\quad Our results show the superiority of the proposed configuration. The methods given in this paper bridge the gap between the chaotic communication theoretical research with the practical communication system configuration and they contribute to related fields. The remaining of the paper is organized as follows. The radio-wave communication system configuration using chaos is presented in Sec.2. The system working principle of the proposed method is reported in Sec.3. In Sec.4, the decoding threshold is introduced. Simulation and test results are reported in Sec.5, verifying the validity and superiority of the proposed method. Finally, some concluding remarks are given in Sec.6.
\section{Radio-wave Communication System Configuration}
In digital communication systems, the time-domain waveform of the transmitted information is treated as a rectangular pulse, which is difficult to transmit because of the broad band requirement. The pulse-shaping filter is used to compress the input signal bandwidth, and to convert the information bit into a baseband signal waveform suitable for the channel transmission. In this paper, a chaotic pulse-shaping filter replace the conventional pulse-shaping filter to generate the baseband signal. The baseband signal spectrum is moved to the bandwidth range by linear modulation so that it is fit for the radio wave channel transmission. Figure. \ref{fig.1} shows the radio wave communication system block diagram, where \emph{G}$_{T}$($\omega$), \emph{C}($\omega$), and \emph{G}$_{T}$($\omega$) represent the pulse shaping filter, transmission channel and the matched filter, respectively; the modulated carrier is used to modulate the baseband signal \emph{u}(\emph{t}) to obtain the frequency band signal \emph{m}(\emph{t}), and the coherent carrier are used to demodulate the received signal \emph{r}(\emph{t}) at the receiver to obtain demodulation signal \emph{d}(\emph{t}), respectively; the decoder is used to decode \emph{\^s}$_{m}$ from the matched filter output signal \emph{y}(\emph{t}); \emph{s}$_{m}$ and \emph{\^s}$_{m}$ represent the transmitted symbol and the received symbol, respectively.
\section{System Working Principle}
From Fig. \ref{fig.1}, we recognize that the system configuration of wireless communication is precisely the conventional radio wave communication system configuration. Therefore, we simply replace the pulse-shaping filter with the chaotic one and also replace the matched filter accordingly. In the following, the principle of radio wave communication system using the properties of chaos is introduced.

\begin{figure}[H]
\centering
\includegraphics[width=3.5in,angle=0]{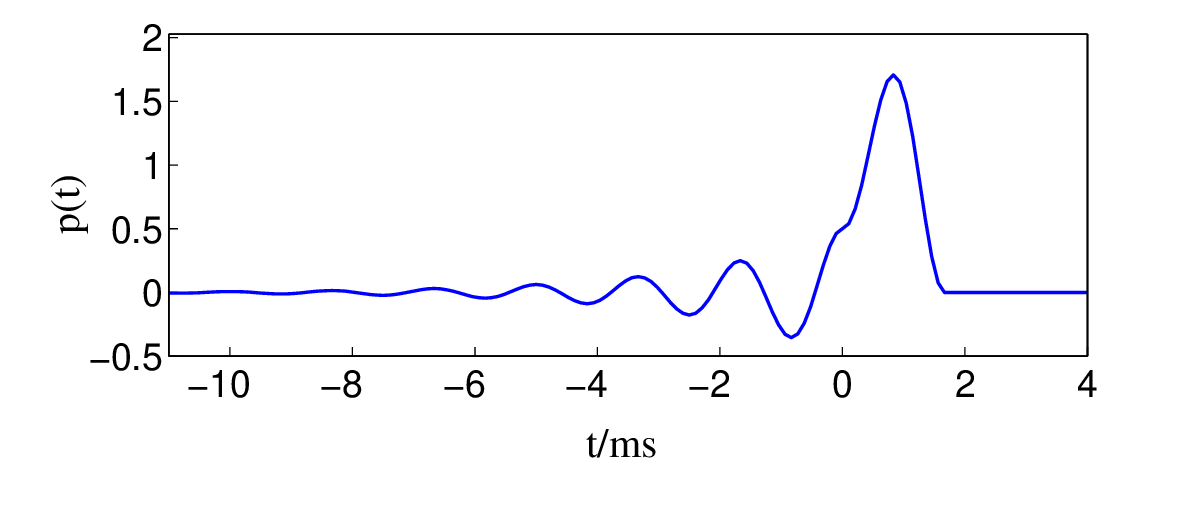}
\caption{\label{fig.2} Time-domain waveform of the pulse shaping filter \emph{p}(\emph{t}), where the base frequency is $f=0.6 kHz$ and the sampling frequency is $f_{s}=9.6 kHz$.}
\end{figure}

In our proposed radio-wave communication system, the pulse-shaping filter is shown in Eq. (\ref{Eq1}) and its time domain waveform is shown in Fig. \ref{fig.2}. In fact, the pulse-shaping filter given by Eq. (\ref{Eq1}) is the basis function of the hybrid system \cite{Corron2016Analytically}, which is used to encode source code stream (information) by using convolution the base function with the bipolar information bit. By this way, the information is encoded in the chaotic baseband signal.
\begin{flalign}\label{Eq1}
&p\left( t \right) = \left\{ \begin{array}{l}
 \left( {1 - {e^{ - \frac{\beta }{f}}}} \right){e^{\beta t}}\left( {\cos \left( {\omega t} \right) - \frac{\beta }{\omega }\sin \left( {\omega t} \right)} \right),\quad t < 0 \\
 1 - {e^{\beta \left( {t - \frac{1}{f}} \right)}}\left( {\cos \left( {\omega t} \right) - \frac{\beta }{\omega }\sin \left( {\omega t} \right)} \right),\quad 0 \le t < \frac{1}{f} \\
 0,\quad t \ge \frac{1}{f} \\
 \end{array} \right.,&
\end{flalign}
where $\omega=2\pi\cdot f$, $\beta=f\cdot\ln2$ are parameters, and \emph{f} is the base frequency of the chaotic signal.
The pulse shaping filter output, \emph{u}(\emph{t}), is given by
\begin{equation}\label{Eq2}
u(t) = \sum\limits_{m =  - \infty }^\infty  {{s_m} \cdot p(t - \frac{m}{f})s(t) = {s_{\left[ {t \cdot f} \right]}}} ,
\end{equation}
where $s_{m}=±1$, $\left[ {t \cdot f} \right]$ is the closest integer in the downward direction.
Given a symbol sequence $[1-11-11, 11-111, -111-1-1, 1-1-1-11, -111-11,-1-1-1-11]$, the pulse shaping filter output is shown in Fig. 3(a), where the red dash line shows the information sequence, the blue solid line is the pulse shaping filter output. Fig. 3(b) shows the spectrum of the pulse shaping filter output. Fig. 3(c) is the three dimensional phase space plot of the states and the information.
\begin{figure}[H]
\centering
\includegraphics[width=3.5in,angle=0]{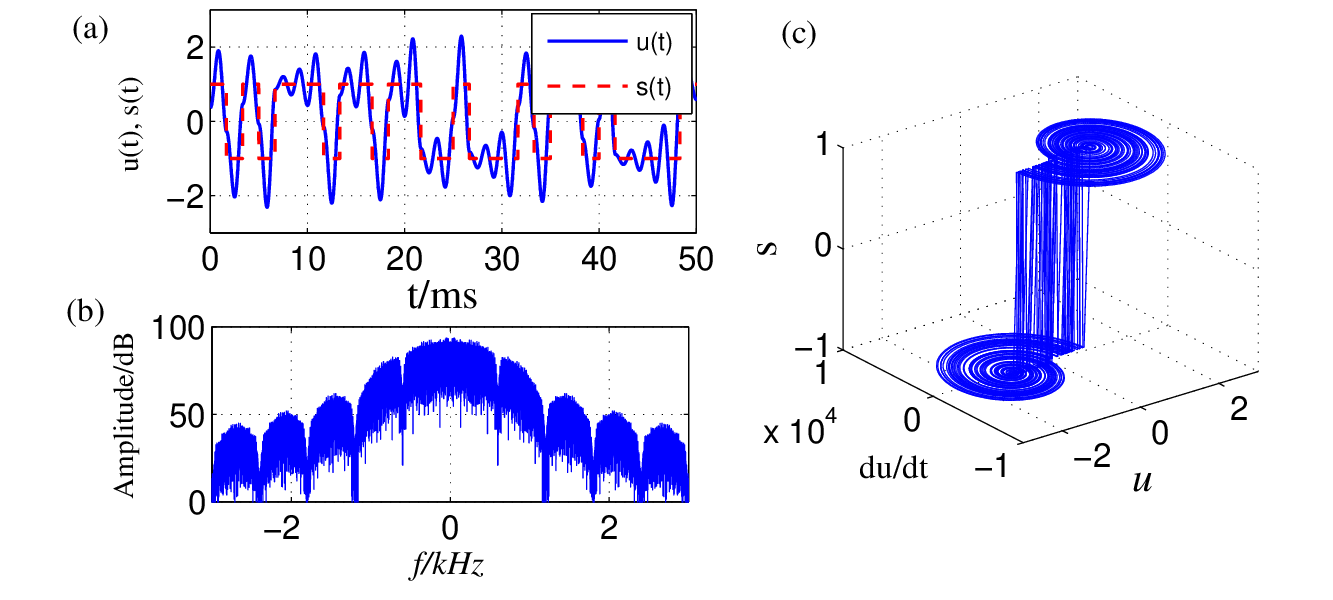}
\caption{\label{fig.3} Chaotic pulse-shaping filter output and its properties with $f=0.6 kHz$ and the sampling frequency is $fs=9.6 kHz$; (a) Filter output and information symbol; (b) The spectrum of the filter output; (c) The three dimensional phase plot of the states and information.}
\end{figure}
As can be seen from Fig. 3(b), the chaotic signal has a rich low-frequency component. In order to meet the bandwidth requirement of the radio wave channel transmission, the spectrum of the baseband signal is shifted to the required channel bandwidth range using linear modulation (referred to as Double Side Band Suppression Carrier, DSB-SC), given by in Eq. (\ref{Eq3}), to obtain the transmitted signal, \emph{m}(\emph{t}).
\begin{equation}\label{Eq3}
m\left( t \right) = u\left( t \right) \cdot \cos \left( {2\pi {f_c}t + \theta } \right),
\end{equation}
where \emph{f}$_{c}$ is the carrier frequency and $\theta$ is the carrier initial phase.

The wireless channel mathematical model \cite{ZhangEYbook} is given by
\begin{equation}\label{Eq4}
h\left( t \right) = \sum\limits_{l = 0}^{L - 1} {{\alpha _l}\delta \left( {t - {\tau _l}} \right)}  + n(t),
\end{equation}
where $\delta(\cdot)$ is Dirac function, $\alpha_{l}$ and $\tau_{l}$ are the attenuation coefficient and the delay time for the $\emph{l}\emph{th}$ multi-path in the channel, respectively, and \emph{n}(\emph{t}) is the Gaussian noise.

At the receiver, the received signal is multiplied by a local carrier synchronized with the modulated carrier, as the following,
\begin{flalign}\label{Eq5}
\begin{array}{l}
 d\left( t \right) = m\left( t \right)\cos \left( {2\pi {f_c}t + \theta } \right) \\
 {\rm{\qquad}}= \frac{1}{2}u\left( t \right) + \frac{1}{2}u\left( t \right)\cos \left( {4\pi {f_c}t + 2\theta } \right). \\
 \end{array}
\end{flalign}

\begin{figure}
\includegraphics[scale=0.45,trim= 0 0 0 0] {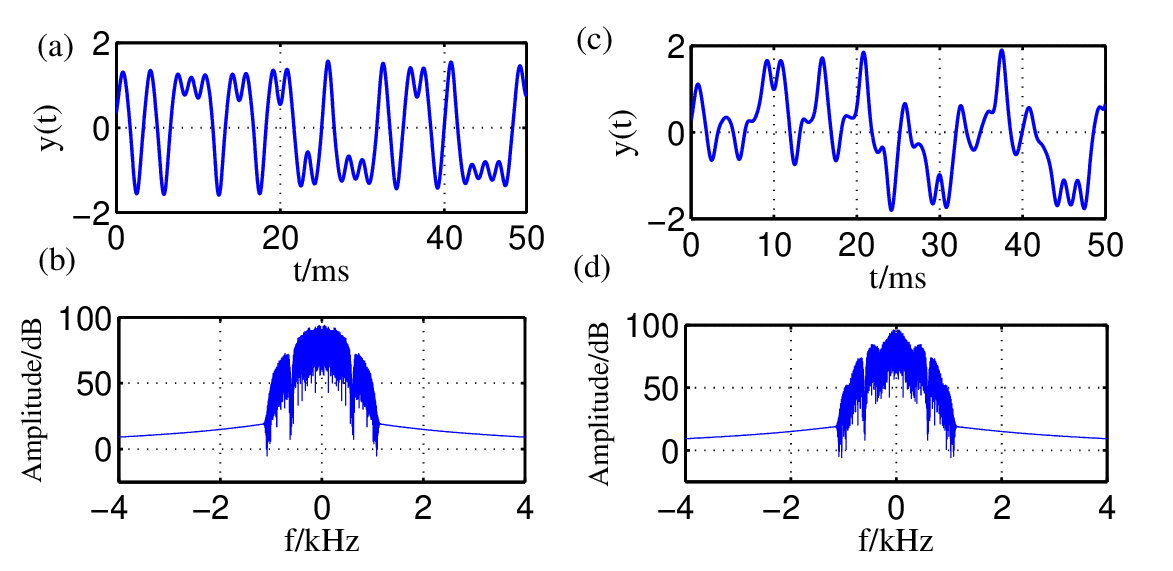}
\caption{\label{fig.4} The output signal waveform of the matched filter and its spectrum, where the base frequency is $f=0.6 kHz$, sampling frequency is $f_{s}=9.6 kHz$, carrier frequency is $f_{c}=1.8 kHz$, and signal to noise ratio $SNR=20 dB$; (a) Filter output for the single path; (b) The spectrum of the signal in (a); (c) The filter output for the two paths; (d) The spectrum of the signal in (c).}
\end{figure}

The corresponding matched filter, with impulse response given by $g(t)=p(-t)$, is used to maximize the signal-to-noise ratio, as shown by
\begin{equation}\label{Eq6}
y\left( t \right) = g\left( t \right) \ast d\left( t \right),
\end{equation}
where `$\ast$' denotes convolution. The matched filter output signal \emph{y}(\emph{t}) waveform and its spectrum are shown in Fig. \ref{fig.4}. Subplots (a) and (c) in Fig. \ref{fig.4} are the matched filter output in the single-path and the double-path channels, respectively; the subplots (b) and (d) in Fig. \ref{fig.4} are the corresponding spectra of the filter output signal in subplots (a) and (c), respectively. The sampled value using sampling interval as $\emph{T}_{s}/2$ ($\emph{T}_{s}$ is the symbol period) in the decoder gets the maximum signal-to-noise ratio.
\section{The Decode Threshold}
In wireless channel, ISI caused by multi-path propagation is unavoidable. In traditional wireless communication systems the ISI effect is decreased by using complicated channel equalization methods. Due to a property of the chaotic signal, ISI effect can be decreased by using the proper decoding threshold \cite{Yao2017Chaos}. Here, the method is briefly explained as follows.
In our proposed communication system, the signal received by the receiver is given by
\begin{flalign}\label{Eq7}
&\begin{array}{*{20}{c}}
   \begin{array}{l}
 r\left( t \right) = h\left( t \right)*u\left( t \right)+n\left( t \right)\\\qquad=\sum\limits_{l=0}^{L{\rm{-}}1}{{\alpha_l}}\sum\limits_{m=-\infty}^\infty{{s_m}p\left({t-{\tau_l}-\frac{m}{f}}\right)}+n\left(t\right), \\
 \end{array}  \\
\end{array}&
\end{flalign}
where \emph{n}(\emph{t}) is Gaussian noise. Then the matched filter output is given by

\begin{flalign}\label{Eq8}
&\begin{array}{l}
 y\left( t \right) = g\left( t \right)*r\left( t \right) = \int_{{\rm{ - }}\infty }^\infty  {p\left( { - \tau } \right)r\left( {t - \tau } \right)d\tau }  \\
  \qquad= \sum\limits_{l = 0}^{L{\rm{ - }}1} {{\alpha _l}} \sum\limits_{m =  - \infty }^\infty  {{s_m}\left( {\int_{ - \infty }^\infty  {p\left( \tau  \right)p\left( {\tau  - t + {\tau _l} + \frac{m}{f}} \right)d\tau } } \right)}  \\
  \qquad=+ \int_{ - \infty }^\infty  {p\left( {\tau  - t} \right)n\left( \tau  \right)d\tau } . \\
 \end{array}&
\end{flalign}

Sampling \emph{y}(\emph{t}) by using frequency \emph{f}, yields
\begin{flalign}\label{Eq9}
&\begin{array}{l}
{y_n}=\sum\limits_{l = 0}^{L - 1} {{\alpha _l}} \sum\limits_{m =  - \infty }^\infty  {{s_m}\int_{ - \infty }^\infty  {p\left( \tau  \right)p\left( {\tau  + {\tau _l} + \frac{{m - n}}{f}} \right)d\tau } }  \\
  \qquad+ \int_{ - \infty }^\infty  {P\left( {\tau  - \frac{n}{f}} \right)n\left( \tau  \right)d\tau }  \\
 \quad = \sum\limits_{l = 0}^{L - 1} {{\alpha _l}\sum\limits_{m =  - \infty }^\infty  {{s_m}{C_{l,m - n}}} }  + W \\
 \quad = \sum\limits_{l = 0}^{L - 1} {{s_n}{C_{l,0}}}  + \sum\limits_{l = 0}^{L - 1} {\sum\limits_{\scriptstyle m =  - \infty  \hfill \atop
  \scriptstyle m \ne n \hfill}^\infty  {{s_m}{C_{l,m - n}}} }  + W \\
  \quad= {s_n}P + I + W \\
 \end{array},&
\end{flalign}
where $\emph{C}_{\emph{l},\emph{j}}$ is the sampled value of the \emph{l}\emph{th} multi-path at the \emph{j}\emph{th} instant, \emph{P} is the sum of multi-path power for $\emph{s}_\emph{n}$, \emph{I} is the interference from other symbols in multi-path, and $\emph{W}$ is the noise effect.

Using the decoding threshold, $\theta=I$ (where $\theta$ is the decoder judgment threshold) removes the inter-symbol interference \cite{Yao2017Chaos}. According to Eq. (\ref{Eq9}), I can be written as
\begin{flalign}\label{Eq10}
\begin{array}{l}
 I{\rm{ = }}\sum\limits_{l = 0}^{L - 1} {\sum\limits_{\scriptstyle i =  - \infty  \hfill \atop
  \scriptstyle i \ne 0 \hfill}^\infty  {{s_{n + i}}{C_{l,i}}} }  \\
 = \sum\limits_{l = 0}^{L - 1} {\sum\limits_{i =  - \infty }^{n - 1} {{s_{n + i}}{C_{l,i}}} }  + \sum\limits_{l = 0}^{L - 1} {\sum\limits_{i = 1}^\infty  {{s_{n + i}}{C_{l,i}}} }  \\
 \end{array}.
\end{flalign}

As can be seen from Eq. (\ref{Eq10}), \emph{I} is decided by both the past symbols transmitted and the future symbols to be transmitted, and as well as by the channel parameters. For the practical communication system the future symbols cannot be predicted, so, the $\emph{I}_{past}$ is used as the suboptimal threshold to decode the information.

\section{The Performance Comparison}

\begin{figure}
\includegraphics[scale=0.6,trim= 0 0 0 0] {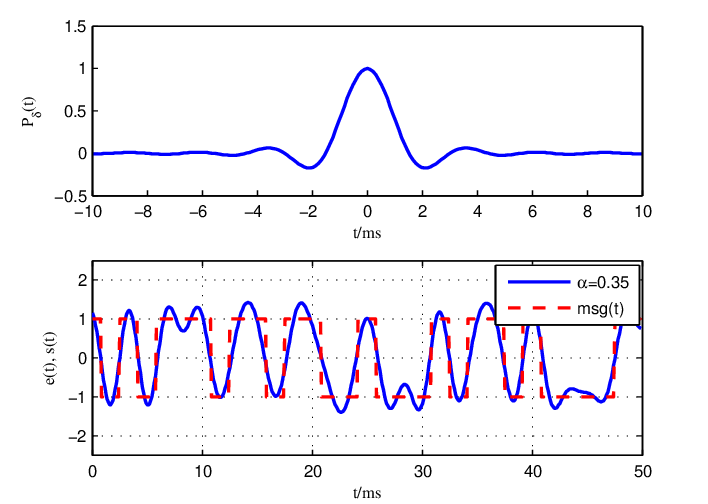}
\caption{\label{fig.5} (a) Root cosine pulse shaping filter waveform with bit rate $r_{b}=600bps$, sampling frequency $f_{s}=9.6 kHz$, and roll-off factor $\gamma=0.35$; (b) Filter output waveform for the symbol sequence [1-11-11, 11-111, -111-1-1, 1-1-1-11, -111-11, -1-1-1-11].}
\end{figure}

\begin{figure}
\includegraphics[scale=0.5,trim= 0 0 0 0] {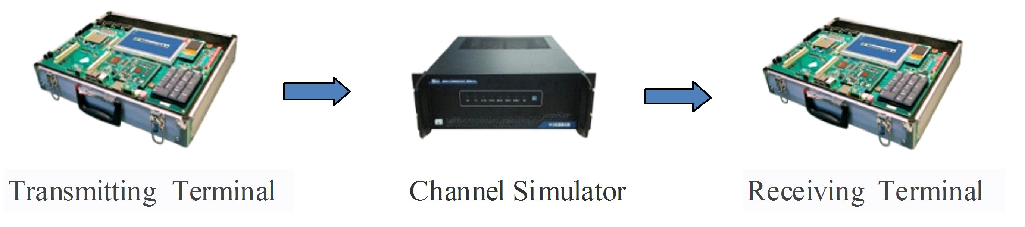}
\caption{\label{fig.6} Chaotic radio wave communication system experimental configuration.}
\end{figure}

In this section, the conventional radio wave communication system is compared with the proposed chaotic radio wave communication system. As we have mentioned in the introduction, one of the main aims of this work is to use chaos in conventional wireless communication system with minimal change to the system configuration. To do this, we only replace the pulse-shaping filter and the matched filter in the conventional wireless communication system with the chaotic pulse-shaping filter given in Eqs. (\ref{Eq1}) and (\ref{Eq2}), and, at the same time, we replace the matched filter at the receiver with the corresponding match filter given by Eq. (\ref{Eq6}). In the conventional radio wave communication system, the root raised cosine pulse-shaping filter (for BPSK modulation) and its corresponding match filter are used, as given by
\begin{flalign}\label{Eq11}
{P_\delta }\left( t \right) = 4\gamma \frac{{\cos \left( {\frac{{\left( {1 + \gamma } \right)\pi t}}{{{T_s}}}} \right) + \frac{{\sin \left( {\frac{{\left( {1 + \gamma } \right)\pi t}}{{{T_s}}}} \right)}}{{\left( {{{4\gamma t} \mathord{\left/
 {\vphantom {{4\gamma t} {{T_s}}}} \right.
 \kern-\nulldelimiterspace} {{T_s}}}} \right)}}}}{{\pi \sqrt {{T_s}} \left( {1 - {{\left( {{{4\gamma t} \mathord{\left/
 {\vphantom {{4\gamma t} {{T_s}}}} \right.
 \kern-\nulldelimiterspace} {{T_s}}}} \right)}^2}} \right)}},
\end{flalign}
where $\gamma$ is roll-off factor and $\emph{T}_{s}$ is the symbol period. The time domain waveform and the filter output for the symbol sequence [1-11-11, 11-111, -111-1-1, 1-1-1-11, -111-11, -1-1-1-1] are shown in Fig. \ref{fig.5}.
The pulse-shaping filter, Eq. (\ref{Eq11}), is commercially used in radio-wave communication system \cite{SQQbook,FanCXbook}. In the following simulation and experimental results, this traditional system is referred to as BPSK for simplicity, the proposed one is referred to as chaos for simplicity.
In this paper, the experimental setup is given in Fig. \ref{fig.6}, which consists of two Digital Signal Processors (6713 DSK made by Texas Instruments) and a hardware channel simulator to generate the channel physical model, which is widely used for radio wave communication test \cite{fenghuo-web}.
\begin{figure}
\includegraphics[scale=0.7,trim= 0 0 0 0] {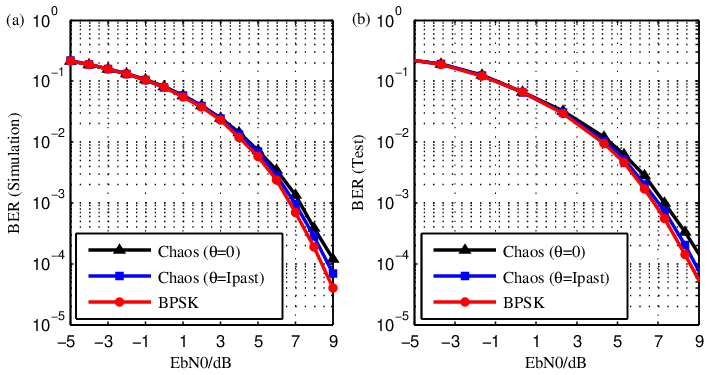}
\caption{\label{fig.7} Simulation and experiment performance comparison of chaotic and BPSK communication systems in single path channel, where the base frequency is $f=0.6 kHz$, sampling frequency is $f_{s}=9.6 kHz$, carrier frequency is $f_{c}=1.8 kHz$, and roll-off factor is $\gamma=0.35$.}
\end{figure}
\begin{figure}
\includegraphics[scale=0.7,trim= 0 0 0 0] {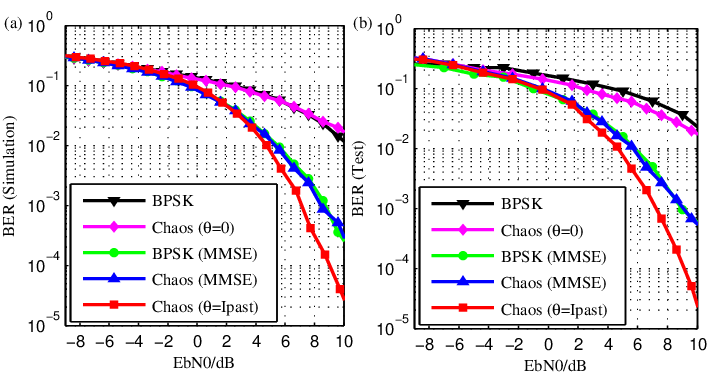}
\caption{\label{fig.8} Simulation and test performance comparison of chaotic and BPSK communication systems in two-path channel, where the base frequency is $f=0.6 kHz$, sampling frequency is $f_{s}=9.6 kHz$, carrier frequency is $f_{c}=1.8 kHz$, and roll-off factor is $\gamma=0.35$, delay time is $\tau$=[0 $\emph{T}_{s}$]($\emph{T}_{\emph{s}}$ is symbol period), attenuation coefficient is $\alpha_{l}=[1 0.6]$; (a) The simulation BER performance comparison; (b) The test BER performance comparison.}
\end{figure}
The simulation and experimental results are shown in Fig. \ref{fig.7} and Fig. \ref{fig.8}, respectively. Figure \ref{fig.7} shows the bit error rate (BER) comparison under the single path wireless channel. Subplots (a) and (b) in Fig. \ref{fig.7} show the simulation and experimental results, respectively. Because it is the single path channel, no equalization is used in both methods. We can see from Fig. \ref{fig.7} that both simulation and experiment results are consistent, and both indicate that, firstly, the proposed chaotic pulse shaping filter using suboptimal threshold achieves better BER performance than using zero as threshold; secondly, the conventional BPSK has better BER performance than the chaotic one in the single path channel. The reason is as follows. The root raised cosine shaping filter is introduced to eliminate the inter-symbol interference in single path channel. However, in the chaotic wireless communication system, although the receiver can also yield the maximum signal-to-noise ratio through the matched filter, the chaotic pulse-shaping filter itself cannot remove ISI influence in the single-path channel. But we can see from Fig. \ref{fig.7} that the BERs of the proposed method and the conventional one are quite similar in single path channel. Figure \ref{fig.8} gives the two-path channel simulation and experiment BER comparison. The black line with down triangular mark and the green line with dot mark in the subplot (a) represent the simulation BER of the BPSK with and without the channel equalization using the minimum mean square error (MMSE) equalizer \cite{Tu2002Minimum}, respectively. The pink line with diamond mark, the blue line with up triangular mark and the red line with rectangular mark, respectively, represents the simulation BER of the chaotic one using 0 as threshold, at the same time using the MMSE equalization, and using $\emph{I}_{\emph{past}}$ as threshold without any equalization. Subplot (b) in Fig. \ref{fig.8} is the experimental BER corresponding to Fig. 8(a). As can be seen from Fig. \ref{fig.8}, under the multi-path channel, the simulation and the experimental test show that, when using $\emph{I}_{\emph{past}}$ as the threshold, even without equalization, the BER of the chaotic one is lower than that of the BPSK even with MMSE. Chaotic system demonstrates better performance in the multi-path channel. For the other multi-path channel parameters, the results are similar to the results in Fig. \ref{fig.8}.
\section{Conclusions}

In this paper, in order to look for a way to use chaos properties in the radio wave communication system with minimum modification to the conventional communication system, we propose to use chaotic pulse shaping filter and its corresponding matched filter to replace the conventional pulse shaping filter. This new wireless communication system achieves maximum SNR, and the use of a simple suboptimal threshold improves the BER performance in multi-path channel, and improves the system performance in a simple way by removing the complicated channel equalization algorithm, thus simplify the software of the system.

\end{document}